\documentclass[aps,prb,showpacs,twocolumn,floats,epsfig]{revtex4}
\usepackage{amssymb}
\usepackage{amsbsy}
\usepackage{amsmath}
\usepackage{epsfig}

\begin{document}

\title {Theory of tunneling conductance of graphene NIS junctions }

\author{Subhro Bhattacharjee$^{(1)}$, Moitri Maiti$^{(2)}$ and K. Sengupta$^{(2)}$}

\affiliation{$^{(1)}$CCMT, Department of Physics, Indian Institute of Science, Bangalore-560012, India.\\
$^{(2)}$TCMP division, Saha Institute of Nuclear Physics, 1/AF
Bidhannagar, Kolkata-700064, India. }

\date{\today}

\begin{abstract}

We calculate the tunneling conductance of a graphene normal
metal-insulator-superconductor (NIS) junction with a barrier of
thickness $d$ and with an arbitrary voltage $V_0$ applied across the
barrier region. We demonstrate that the tunneling conductance of
such a NIS junction is an oscillatory function of both $d$ and
$V_0$. We also show that the periodicity and amplitude of such
oscillations deviate from their universal values in the thin barrier
limit as obtained in earlier work [Phys. Rev. Lett. {\bf 97}, 217001
(2006)] and become a function of the applied voltage $V_0$. Our
results reproduces the earlier results on tunneling conductance of
such junctions in the thin [Phys. Rev. Lett. {\bf 97}, 217001
(2006)] and zero [Phys. Rev. Lett. {\bf 97}, 067007 (2006)] barrier
limits  as special limiting cases. We discuss experimental relevance
of our results.

\end{abstract}


\maketitle

\section{Introduction}

Graphene, a two-dimensional single layer of graphite, has been
recently fabricated by Novoselov {\it et.\,al.} \cite{nov1}. This
has provided an unique opportunity for experimental observation of
electronic properties of graphene which has attracted theoretical
attention for several decades \cite{oldref}. In graphene, the energy
bands touch the Fermi energy at six discrete points at the edges of
the hexagonal Brillouin zone. Out of these six Fermi points, only
two are inequivalent; they are commonly referred to as $K$ and $K'$
points \cite{ando1}. The quasiparticle excitations about these $K$
and $K'$ points obey linear Dirac-like energy dispersion. The
presence of such Dirac-like quasiparticles is expected to lead to a
number of unusual electronic properties in graphene including
relativistic quantum Hall effect with unusual structure of Hall
plateaus \cite{shar1}. Recently, experimental observation of the
unusual plateau structure of the Hall conductivity has confirmed
this theoretical prediction \cite{nov2}. Further, as suggested in
Ref.\ \onlinecite{geim1}, the presence of such quasiparticles in
graphene provides us with an experimental test bed for Klein
paradox. \cite{klein1}

Another, less obvious but nevertheless interesting, consequence of
the existence Dirac-like quasiparticles can be understood by
studying tunneling conductance of a normal metal-superconductor (NS)
interface of graphene \cite{beenakker1}. Graphene is not a natural
superconductor. However, superconductivity can be induced in a
graphene layer in the presence of a superconducting electrode near
it via proximity effect \cite{volkov1,beenakker1,beenakker2} or by
possible intercalation with dopant molecules \cite{uchoa1}. It has
been recently predicted \cite{beenakker1} that a graphene NS
junction, due to the Dirac-like energy spectrum of its
quasiparticles, can exhibit specular Andreev reflection in contrast
to the usual retro reflection observed in conventional NS junctions
\cite{andreev1,tinkham1}. Such specular Andreev reflection process
leads to qualitatively different tunneling conductance curves
compared to conventional NS junctions \cite{beenakker1}. The effect
of the presence of a thin barrier region of thickness $d \rightarrow
0$ created by applying a large gate voltage $ V_0 \rightarrow
\infty$ ( such that $V_0 d$ is finite) between the normal and the
superconducting region has also been studied in Ref.\
\onlinecite{bhattacharya1}. It has been shown that in this thin
barrier limit, in contrast to all normal
metal-insulator-superconductor (NIS) junctions studied so far, the
tunneling conductance of a graphene NIS junction is an oscillatory
function of the dimensionless barrier strength $\chi = V_0 d /(\hbar
v_F)$, where $v_F$ denotes the Fermi velocity of graphene, with
periodicity $\pi$. Further, it has also been demonstrated that the
tunneling conductance reaches its maxima of $2G_0$ for $ \chi =
(n+1/2)\pi$, where $n$ is an integer. The latter result was also
interpreted in terms of transmission resonance property of the
Dirac-Bogoliubov quasiparticles \cite{nov2}. However, no such
studies have been undertaken for NIS junctions with barriers of
arbitrary thickness $d$ and barrier potential $V_0$.

In this work, we extend the analysis of Ref.\
\onlinecite{bhattacharya1} and calculate the tunneling conductance
of a graphene NIS junction with a barrier of thickness $d$ and with
an arbitrary voltage $V_0$ applied across the barrier region. The
main results of our work are the following. First, we show that the
oscillatory behavior of the tunneling conductance is not a property
of the thin barrier limit, but persists for arbitrary barrier width
$d$ and applied gate voltage $V_0$, as long as $d \ll \xi$, where
$\xi$ is the coherence length of the superconductor. Second, we
demonstrate that the periodicity and amplitude of these oscillations
deviate from their values in the thin barrier limit and becomes a
function of the applied voltage $V_0$. We point out that the
barriers which can be realistically achieved in current experimental
setups \cite{nov2} do not necessarily fall in the thin barrier
regime which necessitates a detailed study of arbitrary barriers as
undertaken here. Finally, we show that our analysis correctly
reproduces the tunneling conductance for both zero barrier
\cite{beenakker1} and thin barrier \cite{bhattacharya1} as limiting
cases.

The organization of the rest of the paper is as follows. In Sec.\
\ref{an1}, we develop the theory of tunneling conductance for a
barrier of thickness $d \ll \xi$ and with a voltage $V_0$ applied
across the barrier region and demonstrate that they correctly
reproduce the results of Refs.\ \onlinecite{bhattacharya1} and
\onlinecite{beenakker1} as limiting cases. The results obtained from
this theory is discussed in Sec.\ \ref{results}. Finally, in Sec.\
\ref{experiments}, we discuss possible experiments that can be
performed to test our theory.

\section{Calculation of tunneling conductance}
\label{an1}

\begin{figure}
\rotatebox{0}{
\includegraphics*[width=\linewidth]{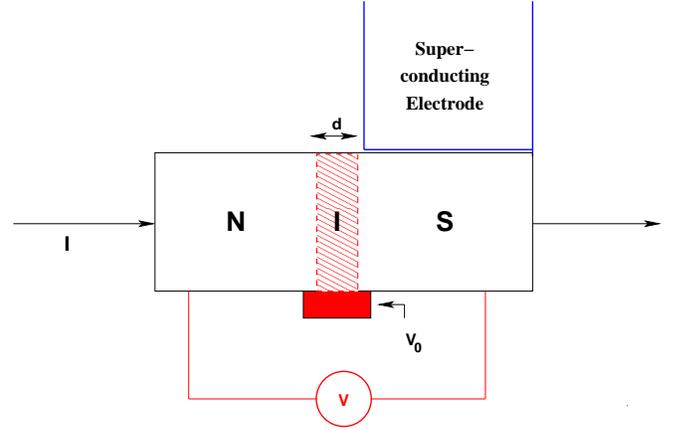}}
\caption{(Color online) A schematic sketch of a graphene NIS
junction. The dashed region sees a variable gate (shown as red
filled region) voltage $V_0$ which creates the barrier. Additional
gate voltage $U_0$, which may be applied on the superconducting
side, and the current source is not shown to avoid clutter. }
\label{figsys}
\end{figure}

Let us consider a NIS junction in a graphene sheet occupying the
$xy$ plane with the normal region occupying $x \le -d$ for all $y$
as shown schematically in Fig.\ \ref{figsys}. The region I, modeled
by a barrier potential $V_0$, extends from $x=-d$ to $x=0$ while the
superconducting region occupies $x\ge 0$. Such a local barrier can
be implemented by either using the electric field effect or local
chemical doping \cite{geim1,nov2}. The region $x \ge 0$ is to be
kept close to an superconducting electrode so that superconductivity
is induced in this region via proximity effect
\cite{volkov1,beenakker1}. In the rest of this work, we shall assume
that the barrier region has sharp edges on both sides. This
condition requires that $d \ll \lambda = 2\pi/k_F$, where $k_F$ and
$\lambda$ are Fermi wave-vector and wavelength for graphene, and can
be realistically created in experiments \cite{geim1}. The NIS
junction can then be described by the Dirac-Bogoliubov-de Gennes
(DBdG) equations \cite{beenakker1}
\begin{eqnarray}
&&\left(\begin{array}{cc}
    {\mathcal H}_{a}-E_F + U({\bf r}) & \Delta ({\bf r}) \\
     \Delta^{\ast}({\bf r}) & E_F - U({\bf r})-{\mathcal H}_{a}
    \end{array}\right) \psi_{a}   = E \psi_{a}. \nonumber\\
\label{bdg1}
\end{eqnarray}
Here, $\psi_a = \left(\psi_{A\,a}, \psi_{B\,a}, \psi_{A\,{\bar
a}}^{\ast}, -\psi_{B\,{\bar a}}^{\ast}\right)$ are the $4$ component
wavefunctions for the electron and hole spinors, the index $a$
denote $K$ or $K'$ for electron/holes near $K$ and $K'$ points,
${\bar a}$ takes values $K'(K)$ for $a=K(K')$, $E_F$ denote the
Fermi energy which can be made non-zero either by doping or by
applying a potential to the graphene sheet, $A$ and $B$ denote the
two inequivalent sites in the hexagonal lattice of graphene, and the
Hamiltonian ${\mathcal H}_a$ is given by
\begin{eqnarray}
{\mathcal H}_a &=& -i \hbar v_F \left(\sigma_x \partial_x + {\rm
sgn}(a) \sigma_y
\partial_y \right). \label{bdg2}
\end{eqnarray}
In Eq.\ \ref{bdg2}, $v_F$ denotes the Fermi velocity of the
quasiparticles in graphene and ${\rm sgn}(a)$ takes values $\pm$ for
$a=K(K')$. The pair-potential $\Delta({\bf r})$ in Eq.\ \ref{bdg1}
connects the electron and the hole spinors of opposite Dirac points.
We have modeled the pair-potential as
\begin{eqnarray}
\Delta({\bf r}) = \Delta_0 \exp(i\phi) \theta(x), \label{deltaeq}
\end{eqnarray}
where $\Delta_0$ and $\phi$ are the amplitude and the phase of the
induced superconducting order parameter respectively and $\theta(x)$
denotes the Heaviside step function.

The potential $U({\bf r})$ gives the relative shift of Fermi
energies in normal, insulating and superconducting regions of
graphene and can be modeled as
\begin{eqnarray}
U({\bf r}) = -U_0 \theta(x) + V_0 \theta(-x) \theta(x+d).
\label{poteq}
\end{eqnarray}
The gate potential $U_0$ can be used to tune the Fermi surface
mismatch between the normal and the superconducting regions. Notice
that the mean-field conditions for superconductivity are satisfied
as long as $\Delta_0 \ll (U_0 +E_f)$; thus, in principle, for large
$U_0$ one can have regimes where $\Delta_0 \ge E_f$
\cite{beenakker1}.

Eq.\ \ref{bdg1} can be solved in a straightforward manner to yield
the wavefunction $\psi$ in the normal, insulating and the
superconducting regions. In the normal region, for electron and
holes traveling the $\pm x$ direction with a transverse momentum
$k_y=q$ and energy $\epsilon$, the (unrenormalized) wavefunctions
are given by
\begin{eqnarray}
\psi_N^{e \pm} &=&  \left(1,\pm e^{\pm i \alpha},0,0\right) \exp
\left[i \left(\pm k_{n} x + q y\right) \right], \nonumber\\
\psi_N^{h \pm} &=& \left(0,0,1,\mp e^{\pm i \alpha'}\right) \exp
\left[i \left(\pm k'_{n} x + q y \right)\right],
\nonumber\\
\sin(\alpha) &=& \frac{\hbar v_F q}{\epsilon +E_F},  \quad
\sin(\alpha') = \frac{\hbar v_F q}{\epsilon - E_F}, \label{wavenorm}
\end{eqnarray}
where the wave-vector $k_n (k'_n)$ for the electron (hole)
wavefunctions are given by
\begin{eqnarray}
k_n(k'_n) &=& \sqrt{\left(\frac{\epsilon +(-) E_F}{\hbar
v_F}\right)^2-q^2 }, \label{ehwave}
\end{eqnarray}
and $\alpha (\alpha')$ is the angle of incidence of the electron
(hole). 

In the barrier region, one can similarly obtain
\begin{eqnarray}
\psi_B^{e \pm} &=& \left(1,\pm e^{\pm i \theta},0,0\right) \exp
\left[i\left(\pm k_{b} x + q y \right)\right], \nonumber\\
\psi_B^{h \pm} &=& \left(0,0,1,\mp e^{\pm i \theta'}\right) \exp
\left[i \left(\pm k'_{b} x + q y \right)\right], \label{barrwave}
\end{eqnarray}
for electron and holes moving along $\pm x$. Here the angle of
incidence of the electron(hole) $\theta(\theta')$ and the wavevector
$k_b(k'_b)$ are given by is
\begin{eqnarray}
\sin\left[\theta(\theta')\right] &=& \hbar v_F q/\left[\epsilon
+(-)(E_F-V_0)\right], \nonumber\\
k_b(k'_b) &=& \sqrt{\left(\frac{\epsilon +(-) (E_F-V_0)}{\hbar
v_F}\right)^2 -q^2}. \label{barrwave2}
\end{eqnarray}
Note that Eq.\ \ref{barrwave} ceases to be the solution of the Dirac
equation (Eq.\ \ref{bdg1}) when $E_F=V_0$ and $\epsilon=0$. For
these parameter values, Eq.\ \ref{bdg1} in the barrier region
becomes $ {\mathcal H}_a \psi_B =0$ which do not have purely
oscillatory solutions. For the rest of this work, we shall restrict
ourselves to the regime $V_0 > E_F$.

In the superconducting region, the BdG quasiparticles are mixtures
of electron and holes. Consequently, the wavefunctions of the BdG
quasiparticles moving along $\pm x$ with transverse momenta $q$ and
energy $\epsilon$, for $(U_0+E_F) \gg \Delta_0, \epsilon$, has the
form
\begin{eqnarray}
\psi_S^{\pm} &=& \left( e^{\mp i \beta}, \mp e^{\pm i \left(\gamma
-\beta\right)}, e^{-i\phi},\mp e^{i\left(\pm \gamma -\phi \right)}
\right) \nonumber\\
&& \times \exp\left[ i\left(\pm k_s x +q y\right) - \kappa x\right],
\nonumber\\
\sin(\gamma) &=& \hbar v_F q/(E_F + U_0),
 \label{supwave}
\end{eqnarray}
where $\gamma$ is the angle of incidence for the quasiparticles.
Here the wavevector $k_s$ and the localization length $\kappa^{-1}$
can be expressed as a function of the energy $\epsilon$ and the
transverse momenta $q$ as
\begin{eqnarray}
k_s &=& \sqrt{\left[\left(U_0+E_F\right)/\hbar v_F\right]^2 -q^2},
\nonumber\\
\kappa^{-1} &=&  \frac{(\hbar v_F)^2 k_s}{\left[(U_0+E_F) \Delta_0
\sin(\beta)\right]}, \label{local}
\end{eqnarray}
where $\beta$ is given by
\begin{eqnarray}
\beta &=& \cos^{-1} \left(\epsilon/\Delta_0\right) \quad {\rm if}
\left|\epsilon\right| < \Delta_0 ,\nonumber\\
&=& -i \cosh^{-1} \left(\epsilon/\Delta_0\right) \quad {\rm if}
\left|\epsilon\right| > \Delta_0.\label{betaeq}
\end{eqnarray}
Note that for $\left|\epsilon\right| > \Delta_0$, $\kappa$ becomes
imaginary and the quasiparticles can propagate in the bulk of the
superconductor.

Next we note that for the Andreev process to take place, the angles
$\theta$, $\theta'$ and $\alpha'$ must all be less than
$90^{\circ}$. This sets the limit of maximum angle of incidence
$\alpha$. Using Eqns.\ \ref{wavenorm} and \ref{barrwave2}, one finds
that the critical angle of incidence is
\begin{eqnarray}
\alpha_c &=& \alpha_c^{(1)} \theta(V_0-2E_F) + \alpha_c^{(2)}
\theta(2E_F-V_0)  \nonumber\\
\alpha_c^{(1)} &=&
\arcsin\left[\left|\epsilon -
E_F\right|/\left(\epsilon + E_F\right)\right], \nonumber\\
\alpha_c^{(2)} &=& \arcsin\left[\left|\epsilon -
|E_F-V_0|\right|/\left(\epsilon + E_F\right)\right]. \label{criti1}
\end{eqnarray}
Note that in the thin or zero barrier limits treated in Refs.\
\onlinecite{bhattacharya1} and \onlinecite{beenakker1},
$\alpha_c=\alpha_c^{(1)}$ for all parameter regimes.

\begin{figure}
\rotatebox{0}{
\includegraphics*[width=\linewidth]{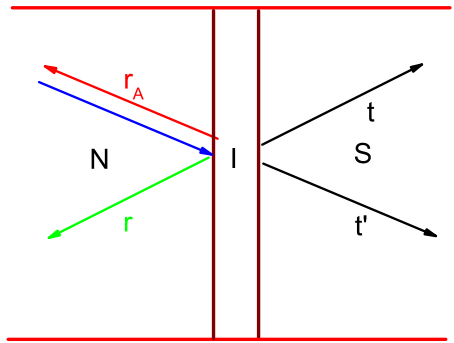}}
\caption{(Color online) A schematic sketch of normal reflection
($r$), Andreev reflection ($r_A$) and transmission processes ($t$
and $t'$) at a graphene NIS junction. Note that in this schematic
picture, we have chosen $r_A$ to denote a retro Andreev reflection
for illustration purpose. In practice, as discussed in the text,
$r_A$ takes into account possibilities of both retro and specular
Andreev reflections. The electron and hole wavefunctions inside the
barrier region is not sketched to avoid clutter.} \label{figprocess}
\end{figure}

Let us now consider a electron-like quasiparticle incident on the
barrier from the normal side with an energy $\epsilon$ and
transverse momentum $q$. The basic process of ordinary and Andreev
reflection that can take place at the interface is schematically
sketched in Fig.\ \ref{figprocess}. As noted in Ref.\
\onlinecite{beenakker1}, in contrast to conventional NIS junction,
graphene junctions allow for both retro and specular Andreev
reflections. The former dominates when $\epsilon, \Delta_0 \ll E_F$
so that $ \alpha = -\alpha'$ (Eq.\ \ref{wavenorm}) while that latter
prevails when $E_F \ll \epsilon, \Delta_0$ with $\alpha = \alpha'$.
Note that in Fig.\ \ref{figprocess}, we have chosen $r_A$ to denote
a retro Andreev reflection for illustration purposes. In practice,
$r_A$ includes both retro and specular Andreev reflections. In what
follows, we shall denote the total probability amplitude of Andreev
reflection as $r_A$ which takes into account possibilities of both
retro and specular Andreev reflections.

The wave functions in the normal, insulating and superconducting
regions, taking into account both Andreev and normal reflection
processes, can then be written as \cite{tinkham1}
\begin{eqnarray}
\Psi_N &=& \psi_N^{e +}+r \psi_N^{e -} + r_A \psi_N^{h -}, \quad
\Psi_S = t \psi_S^{+}+ t' \psi_S^{-}, \nonumber\\
\Psi_B &=& p \psi_B^{e +}+q \psi_B^{e -} + m \psi_B^{h +} + n
\psi_N^{h -}, \label{wave2}
\end{eqnarray}
where $ r$ and $r_A$ are the amplitudes of normal and Andreev
reflections respectively, $t$ and $t'$ are the amplitudes of
electron-like and hole-like quasiparticles in the superconducting
region and $p$, $q$, $m$ and $n$ are the amplitudes of electron and
holes in the barrier. These wavefunctions must satisfy the
appropriate boundary conditions:
\begin{eqnarray}
\Psi_N |_{x=-d} &=& \Psi_B |_{x=-d},  \quad  \Psi_B |_{x=0} = \Psi_S
|_{x=0}. \label{bc1}
\end{eqnarray}
Notice that these boundary conditions, in contrast their
counterparts in standard NIS interfaces, do not impose any
constraint on derivative of the wavefunctions at the boundary. These
boundary conditions yield
\begin{eqnarray}
e^{-ik_n d}+r e^{ik_n d} &=& p e^{-ik_{b}d}+ q e^{ik_{b}d},\nonumber\\
e^{i\alpha} e^{-ik_nd} - r e^{-i\alpha} e^{ik_nd} &=& -p
e^{i(\theta-k_{b}d)}+q e^{-i(\theta-k_{b} d)}, \nonumber\\
r_A e^{ik'_{n}d} &=& m e^{-ik'_{b}d}+n e^{ik'_{b}d},\nonumber\\
r_A e^{-i\alpha'} e^{ik'_n d} &=& -m e^{i(\theta'- k'_{b}d)} +n
e^{-i(\theta' -k'_{b})},\nonumber\\
p+q &=& t e^{-i \beta}+ t' e^{i\beta},\nonumber\\
-p e^{i\theta}+ q e^{-i\theta} &=& -t e^{i(\gamma-\beta)}+t'
e^{-i(\gamma -\beta)}, \nonumber\\
m+n &=& (t+t') e^{-i\phi}.\nonumber\\
-m e^{i\theta'}+n e^{-i\theta'} &=& -t e^{i(\gamma-\phi)}+t'
e^{-i(\gamma+\phi)}. \label{bc2}
\end{eqnarray}

Using the boundary conditions (Eq.\ \ref{bc2}), one can now solve
for the coefficients $r$, $r_A$, $t$ and $t'$ in Eq.\ \ref{wave2}.
After some straightforward but cumbersome algebra, we find that
\begin{eqnarray}
r &=& e^{-2i k_n d} \frac{\mathcal N}{\mathcal D}, \label{req} \\
{\mathcal N} &=& \left[e^{i\alpha}\cos(k_{b}d+\theta)-i \sin(k_{b}d)\right]\nonumber\\
&& -\rho[\cos(k_{b}d-\theta)-i\ e^{i\alpha} \sin(k_{b}d)], \label{neq}\\
{\mathcal D} &=& \left [e^{-i\alpha}\cos(k_{b}d+\theta)
+i\sin(k_{b}d)\right]\nonumber\\
&& + \rho\left[\cos(k_{b}d-\theta)+ie^{-i\alpha}\sin(k_{b}d)\right],  \label{deq} \\
t'&=&{\frac{e^{-ik_n d}}{\cos(\theta) [\Gamma
e^{-i\beta}+e^{i\beta}]}} \Big([\cos(k_{b}d-\theta)-ie^{i\alpha} \sin(k_{b}d)]\nonumber\\
&& +r e^{ik_n d}[\cos(k_{b}d-\theta)+ie^{-i\alpha}\sin(k_{b}d)] \Big),\\
t &=& \Gamma t',\\
r_A &=& \frac{ t (\Gamma + 1)e^{ik'_n d}
\cos(\theta')e^{-i\phi}}{\cos(k'_{b}d-\theta')-ie^{-i\alpha'}
\sin(k'_{b}d)}, \label{raeq}
\end{eqnarray}
where the parameters $\Gamma$ and $\rho$ can be expressed in terms
of $\gamma$, $\beta$, $\theta$, $\theta'$, $\alpha$, and $\alpha'$
(Eqs.\ \ref{wavenorm}, \ref{barrwave2}, \ref{supwave}, and
\ref{betaeq}) as
\begin{eqnarray}
\rho &=& \frac{-\Gamma e^{i(\gamma-\beta)}+e^{-i(\gamma-\beta)}}
{\Gamma e^{-i\beta}+e^{i\beta}},\\
\Gamma &=& \frac{e^{-i\gamma}-\eta}{e^{i\gamma}+\eta},\\
\eta &=& \frac{e^{-i\alpha'} \cos(k'_{b}d+\theta')-i\sin(k'_{b}d)}
{\cos(k'_{b}d-\theta')-ie^{-i\alpha'} \sin(k'_{b}d)}. \label{qt1}
\end{eqnarray}
The tunneling conductance of the NIS junction can now be expressed
in terms of $r$ and $r_A$ by \cite{tinkham1}
\begin{eqnarray}
\frac{G(eV)}{G_0(eV)} &=&  \int_0^{\alpha_c}
\left(1-\left|r\right|^2 + \left|r_A\right|^2
\frac{\cos(\alpha')}{\cos(\alpha)} \right) \cos(\alpha) \, d\alpha,
\nonumber \\ \label{tc1}
\end{eqnarray}
where $G_0 = 4e^2 N(eV)/h$ is the ballistic conductance of metallic
graphene, $eV$ denotes the bias voltage, and $N(\epsilon)= (E_F
+\epsilon)w/(\pi \hbar v_F)$ denotes the number of available
channels for a graphene sample of width $w$. For $eV \ll E_F$, $G_0$
is a constant. Eq.\ \ref{tc1} can be evaluated numerically to yield
the tunneling conductance of the NIS junction for arbitrary
parameter values. We note at the outset, that $G=0$ when
$\alpha_c=0$. This occurs in two situations. First, when $eV=E_F$
and $V_0 \ge 2E_F$ so that $\alpha_c=\alpha_c^{(1)}$ vanishes. For
this situation to arise, $E_F +U_0> \Delta > E_F$ which means that
$U_0$ has to be finite. Second, $\alpha_c=\alpha_c^{(2)}=0$ when
$eV=0$ and $E_F=V_0$, so that the zero-bias conductance vanishes
when the barrier potential matches the Fermi energy of the normal
side \cite{comment1}.

We now make contact with the results of Ref.\
\onlinecite{bhattacharya1} in the thin barrier limit. We note that
since there are no condition on the derivatives of wavefunctions in
graphene NIS junctions, the standard delta function potential
approximation for thin barrier \cite{tinkham1} can not be taken the
outset, but has to be taken at the end of the calculation. This
limit is defined as $d/\lambda \rightarrow 0$ and $V_0/E_F
\rightarrow \infty$ such that the dimensionless barrier strength
\begin{eqnarray}
\chi &=&  V_0 d/\hbar v_F = 2\pi \left(\frac{V_0}{E_F}\right) \left(
\frac{d}{\lambda}\right) \label{barstr}
\end{eqnarray}
remains finite. In this limit, as can be seen from Eqs.\
\ref{wavenorm}, \ref{barrwave2} and \ref{supwave}, $\theta, \theta',
k_n d, k'_n d \rightarrow 0$ and $k_b d, k'_b d \rightarrow \chi$ so
that the expressions for $\Gamma$, $\rho$ and $\eta$ (Eq.\
\ref{qt1})
\begin{eqnarray}
\Gamma^{\rm tb} &=& \frac{e^{-i\gamma} -\eta^{\rm tb}}{e^{i\gamma}
+\eta^{\rm tb}}, \quad \eta^{\rm tb}
 = \frac{e^{- i \alpha'} \cos(\chi)  - i \sin(\chi)}{
\cos(\chi) - i e^{-i \alpha'} \sin(\chi)}, \nonumber\\
\rho^{\rm tb} &=& \frac{e^{-i(\gamma - \beta)} - \Gamma^{\rm tb}
e^{i(\gamma - \beta)}}{\Gamma^{\rm tb} e^{-i\beta} + e^{i \beta}}.
\label{coefftb1}
\end{eqnarray}
where the superscript "${\rm tb}$" denotes thin barrier. Using the
above-mentioned relations, we also obtain
\begin{eqnarray}
r^{\rm tb} &=& \frac{\cos(\chi) \left(e^{i \alpha}-\rho^{\rm tb}
\right) - i \sin(\chi)\left(1-\rho^{\rm tb} e^{i \alpha}
\right)}{\cos(\chi) \left(e^{-i \alpha}+\rho^{\rm tb}\right) + i
\sin(\chi)\left(1+\rho^{\rm tb} e^{-i \alpha}
\right)},\nonumber\\
 t^{'{\rm tb}} &=& \frac{\cos(\chi) \left(1+r^{\rm tb}\right) - i \sin(\chi)\left(e^{i
\alpha}-r^{\rm tb} e^{-i \alpha}\right)}{\Gamma e^{-i\beta} + e^{i
\beta}}, \nonumber\\
t^{\rm tb} &=&  \Gamma t^{'{\rm tb}},\nonumber\\
r_A^{\rm tb} &=& \frac{t'^{\rm tb} \left(\Gamma+1\right) e^{-i
\phi}}{\cos(\chi) - i e^{-i \alpha'} \sin(\chi)}. \label{coefftb2}
\end{eqnarray}
Eqs. \ref{coefftb1} and \ref{coefftb2} are precisely the result
obtained in Ref.\ \onlinecite{bhattacharya1} for the tunneling
conductance of a thin graphene NIS junction. The result obtained in
Ref.\ \onlinecite{beenakker1} can be now easily obtained from Eqs.\
\ref{coefftb1} and \ref{coefftb2} by substituting $\chi=0$ in these
equations, as also noted in Ref.\ \onlinecite{bhattacharya1}.

\section{Results}
\label{results}

\subsection {Qualitative Discussions}

In this section, we shall analyze the formulae for tunneling
conductance obtained in Sec.\ \ref{an1}. First we aim to obtain a
qualitative understanding of the behavior of the tunneling
conductance for finite barrier strength. To this end, we note from
Eq.\ \ref{tc1} that the maxima of the tunneling conductance must
occur where $|r|^2$ is minimum. In fact, if $|r|^2=0$ for all
transverse momenta, the tunneling conductance reaches its value
$2G_0$. Therefore we shall first try to analyze the expression of
$r$ (Eq.\ \ref{req}) for subgap voltages and when the Fermi surfaces
of the normal and superconducting sides are aligned with each other
($U_0=0$). In this case, we need $\Delta_0 \ll E_F$. So for subgap
tunneling conductance, we have $\epsilon \le \Delta_0 \ll E_F$. In
this limit, $\alpha \simeq -\alpha' \simeq \gamma$ (Eqs.
\ref{wavenorm} and \ref{supwave}), $k_b \simeq k_b'$, and $\theta
\simeq -\theta'$ (Eq. \ref{barrwave2}). Using these, one can write
\begin{eqnarray}
\eta &=& \frac{e^{i\alpha} \cos(k_b d - \theta) - i\sin(k_b
d)}{\cos(k_b d + \theta) - i e^{i \alpha} \sin(k_b d)}, \\
\rho &=& \frac{ \eta \cos(\alpha -\beta) + i
\sin(\beta)}{\cos(\alpha + \beta) + i \eta \sin(\beta)}.
\label{rhoeta}
\end{eqnarray}
Substituting Eq.\ \ref{rhoeta} in the expression of ${\mathcal N}$,
we find that the numerator of the reflection amplitude $r$ becomes
(Eqs.\ \ref{req} and \ref{neq})
\begin{eqnarray}
{\mathcal N} &=& \frac{e^{i \alpha}}{D_0} \Bigg[ -4 \sin(\alpha)
\sin(\beta) \cos(k_b d -\theta) \nonumber\\
&& \times \Big[- i\cos(\alpha) \sin(k_b d)\nonumber\\
&& + (\cos(k_b d - \theta)+\cos(k_b d + \theta))/2 \Big] \nonumber\\
&& + 2 \left[\cos(k_b d + \theta)-\cos(k_b d - \theta) \right]
\nonumber\\
&& \times \Big [ \cos(\alpha-\beta) \left\{ \cos(\alpha) +
\left[\cos(k_b
d - \theta) \right. \right. \nonumber\\
&& \left. \left.+ \cos(k_b d + \theta) \right]/2 \right\}  +
\sin(k_B
d) \sin(\beta)\Big] \Bigg], \label{rexp} \\
 D_0 &=&  \cos(k_b d + \theta)\cos(\alpha +
\beta) + \sin(k_b d)
\sin(\beta) \nonumber\\
&& + i e^{i \alpha} \left[ \cos(k_b d - \theta) \sin(\beta) -
\sin(k_b d) \cos(\alpha + \beta) \right]. \nonumber\\ \label{denexp}
\end{eqnarray}

\begin{figure}
\rotatebox{-90}{
\includegraphics*[width=6cm]{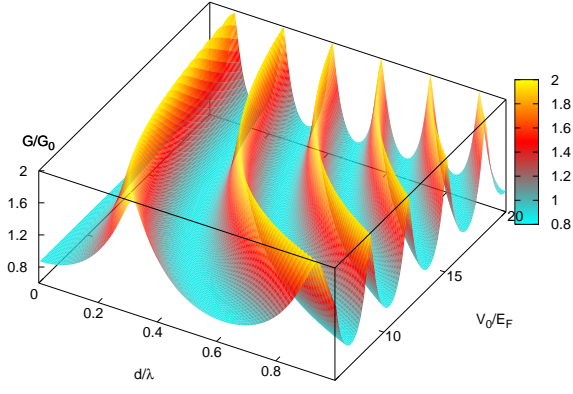}}
\caption{(Color online) Plot of zero-bias tunneling conductance for
$U_0=0$ and $\Delta_0=0.01 E_F$ as a function of gate voltage $V_0$
and barrier thickness $d$. Note that the oscillatory behavior of the
tunneling conductance persists for the entire range of $V_0$ and
$d$.} \label{figm1}
\end{figure}

From the expression of ${\mathcal N}$ (Eq.\ \ref{rexp}), we note the
following features. First, for normal incidence ($\alpha=0$) where $
\theta=\theta'=0$, ${\mathcal N}$ and hence $r$ (Eq.\ \ref{req})
vanishes. Thus the barrier is reflectionless for quasiparticles
which incident normally on the barrier for arbitrary barrier
thickness $d$ and strength of the applied voltage $V_0$. This is a
manifestation of Klein paradox for Dirac-Bogoliubov quasiparticles
\cite{klein1}. However, this feature is not manifested in tunneling
conductance $G$ ( Eq.\ \ref{tc1}) which receives contribution from
all angles of incidence. Second, apart from the above-mentioned
cases, $r$ never vanishes for all angles of incidence $\alpha$ and
arbitrary $eV < \Delta_0$ unless $\theta= \theta'$. Thus the subgap
tunneling conductance is not expected to reach a maximum value of
$2G_0$ as long as the thin barrier limit is not satisfied. However,
in practice, for barriers with $V_0>4E_F$, the difference between
$\theta$ and $\theta'$ turns out to be small for all $q \le k_F$
($\le 0.25$ for $q\le k_F$ and $eV=0$) so that the contribution to
${\mathcal N}$ (Eq.\ \ref{rexp}) from the terms $\sim (\cos(k_b d +
\theta) -\cos(k_b d -\theta))$ becomes negligible. Thus $|r|^2$ can
become quite small for special values of $V_0$ for all $q \le k_F$
so that the maximum value of tunneling conductance can reach close
to $2G_0$. Third, for large $V_0$, for which the contribution of
terms $\sim (\cos(k_b d + \theta) -\cos(k_b d -\theta))$ becomes
negligible, ${\mathcal N}$ and hence $r$ becomes very small when the
applied voltage matches the gap edge ${\it i.e.}$ $\sin (\beta)=0$
(Eq.\ \ref{rexp}). Thus the tunneling conductance curves approaches
close to its maximum value $2G_0$ and becomes independent of the
gate voltage $V_0$ at the gap edge $eV=\Delta_0$ for $\Delta_0 \ll
E_F$, as is also seen for conventional NIS junctions
\cite{tinkham1}. Fourth, in the thin barrier limit, ($V_0/E_F
\rightarrow \infty$ and $d/\lambda \rightarrow 0$), $\theta
\rightarrow 0$ and $k_b d \rightarrow \chi$, so that the
contribution of the terms $\sim (\cos(k_b d + \theta) -\cos(k_b d
-\theta))$ in Eq.\ \ref{rexp} vanishes and one gets
\begin{figure}
\rotatebox{0}{
\includegraphics*[width=\linewidth]{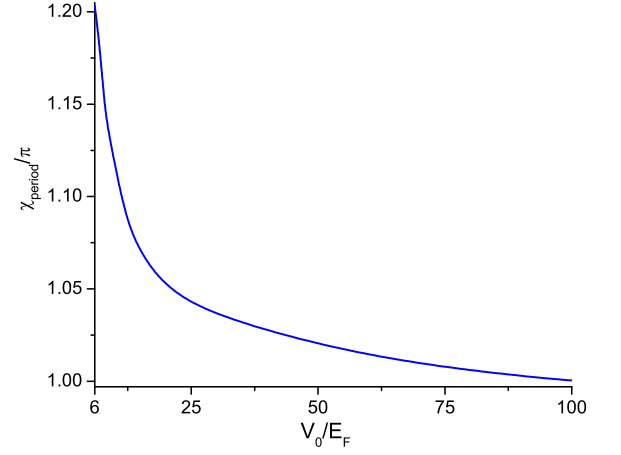}}
\caption{Plot of periodicity $\chi_{\rm period}$ of oscillations of
tunneling conductance as a function of applied gate voltage $V_0$
for $U_0=0$ and $\Delta_0=0.01 E_F$. Note that the periodicity
approaches $\pi$ as the voltage increases since the junction
approaches the thin barrier limit. } \label{figm2}
\end{figure}
\begin{eqnarray}
{\mathcal N}^{tb} &=& \frac{2 \sin(\alpha) [\sin(\chi + \beta)-
\sin(\chi-\beta)] }{D_0^{\rm tb} } \nonumber\\
&& \times \left[ -\cos(\chi) + i \sin(\chi) \cos(\alpha) \right],
\label{ntbeq} \\
D_0^{\rm tb} &=& \cos(\chi) \cos(\alpha + \beta) + \sin(\chi)
\sin(\beta) + i
e^{i \alpha} \nonumber\\
&& \times \left[ \cos(\chi) \sin(\beta) - \sin(\chi) \cos(\alpha +
\beta) \right].
\end{eqnarray}
As noted in Ref.\ \onlinecite{bhattacharya1}, ${\mathcal N}^{tb}$
and hence $r^{\rm tb}$ (Eq.\ \ref{coefftb2}) vanishes at $\chi =
(n+1/2) \pi$ which yields the transmission resonance condition for
NIS junctions in graphene. Fifth, as can seen from Eqs.\ \ref{req}
and \ref{raeq}, both $|r|^2$ and $|r_A|^2$ are periodic functions of
$V_0$ and $d$ since both $k_b$ and $\theta$ depend on $V_0$. Thus
the oscillatory behavior of subgap tunneling conductance as a
function of applied gate voltage $V_0$ or barrier thickness $d$ is a
general feature of graphene NIS junctions with $d \ll \xi$. However,
unlike the thin barrier limit, for an arbitrary NIS junction, $k_b d
= \chi \sqrt{ (E_F/V_0-1)^2 + \hbar^2 v_F^2 q^2 /V_0^2} \neq \chi$,
and $\theta \neq 0$. Thus the period of oscillations of $|r|^2$ and
$|r_A|^2$ will depend on $V_0$ and should deviate from their
universal value $\pi$ in the thin barrier limits. Finally, we note
from Eqs.\ \ref{req}, \ref{tc1} and \ref{ntbeq} that in the thin
barrier limit (and therefore for large $V_0$), the amplitude of
oscillations of the zero-bias conductance for a fixed $V_0$, defined
as $[G_{\rm max}(eV=0;V_0)-G_{\rm min}(eV=0;V_0)]/G_0$, which
depends on the difference of $|r(\chi=(n+1/2)\pi)|^2$ and
$|r(\chi=n\pi)|^2$ becomes independent of $\chi$ or the applied gate
voltage $V_0$.

\begin{figure}
\rotatebox{0}{
\includegraphics*[width=\linewidth]{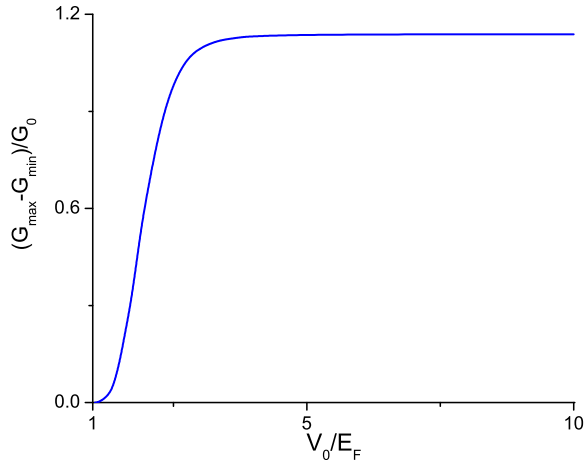}}
\caption{Plot of the amplitude $[G_{\rm max}(eV=0;V_0)-G_{\rm
min}(eV=0;V_0)]/G_0 \equiv (G_{\rm max}-G_{\rm min})/G_0$ of
zero-bias tunneling conductance as a function of the applied gate
voltage $V_0$ for $U_0=0$ and $\Delta_0=0.01 E_F$. Note that $G$
reaches $2G_0$ for $V_0 \ge 4E_F$ where the amplitude become
independent of the applied gate voltage as in the thin barrier limit
and vanishes for $V_0/E_F=1$ as discussed in the text.}
\label{figm3}
\end{figure}

\subsection{Numerical Results}

\begin{figure}
\rotatebox{-90}{
\includegraphics*[width=6cm]{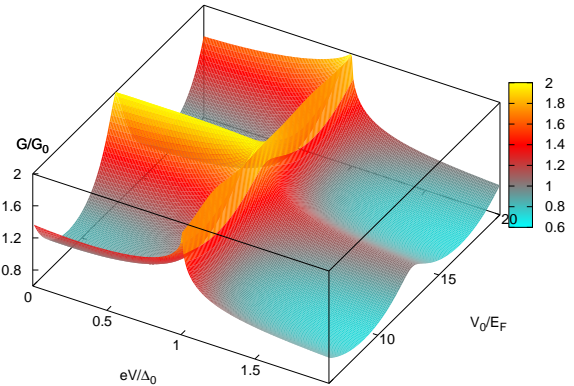}}
\caption{(Color online) Plot of tunneling conductance as a function
of the bias-voltage $eV$ and gate voltage $V_0$ for $d=0.4 \lambda$
and $\Delta_0=0.01 E_F$. Note that for large $V_0$, the tunneling
conductance at $eV= \Delta_0$ is close to $2G_0$ and becomes
independent of $V_0$ (see text for discussion). } \label{figm4}
\end{figure}

The above-mentioned discussion is corroborated by numerical
evaluation of the tunneling conductance as shown in Figs.\
\ref{figm1}, \ref{figm2}, \ref{figm3} and \ref{figm4}. From Fig.\
\ref{figm1}, which plots zero-bias tunneling conductance $G(eV=0)$
as a function of $V_0$ and $d$, we find that $G(eV=0)$ is an
oscillatory function of both $V_0$ and $d$ and reaches close to its
maximum value of $2G_0$ throughout the plotted range of $V_0$ and
$d$. Further, as seen from Fig.\ \ref{figm2}, the periodicity of
these oscillations becomes a function of $V_0$. To measure the
periodicity of these oscillations, the tunneling conductance is
plotted for a fixed $V_0$ as a function of $d$. The periodicity of
the conductance $d_{\rm period}$ is noted down from these plots and
$\chi_{\rm period} = V_0 d_{\rm period}/\hbar v_F$ is computed.
Fig.\ \ref{figm2} clearly shows that $\chi_{\rm period}$ deviate
significantly from their thin barrier value $\pi$ for low enough
$V_0$ and diverges at $V_0 \to E_F$ \cite{comment2}. Fig.\
\ref{figm3} shows the amplitude of oscillations of zero-bias
conductance as a function of $V_0$. We note that maximum of the
zero-bias tunneling conductance $G_{\rm max}(eV=0)$ reaches close to
$2G_0$ for $V_0 \ge V_{0c} \simeq 4E_F$. For $V \ge V_{0c}$, the
amplitude becomes independent of the applied voltage as in the thin
barrier limit, as shown in Fig.\ \ref{figm3}. For $V_0 \to E_F$,
$\alpha_c=\alpha_c^{(2)} \to 0$, so that $G(eV=0) \to 0$ and hence
the amplitude vanishes. Finally, in Fig.\ \ref{figm4}, we plot the
tunneling conductance $G$ as a function of the applied bias-voltage
$eV$ and applied gate voltage $V_0$ for $d=0.4 \lambda$. We find
that, as expected from Eq.\ \ref{ntbeq}, $G$ reaches close to $2G_0$
at the gap edge for all $V_0 \ge 6E_F$. Also, as in the thin barrier
limit, the oscillation amplitudes for the subgap tunneling
conductance is maximum at zero-bias and shrinks to zero at the gap
edge $eV=\Delta_0$, where the tunneling conductance become
independent of the gate voltage.

\begin{figure}
\rotatebox{0}{
\includegraphics*[width=\linewidth]{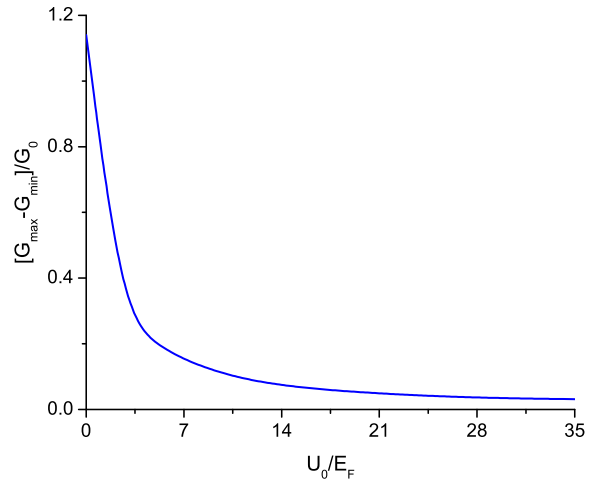}}
\caption{Plot of amplitude of oscillation $(G_{\rm max} -G_{\rm
min})/G_0$ of zero-bias tunneling conductance as a function of
$U_0/E_F$ for $V_0=6 E_F$ and $\Delta_0=0.01 E_F$. The oscillation
amplitudes always decay monotonically with increasing $U_0$
independent of $V_0$.} \label{figm6}
\end{figure}

Next, we consider the case $U_0 \neq 0$, so that $\Delta_0 \simeq
E_F \ll (E_F + U_0)$. In this regime, there is a large mismatch of
Fermi surfaces on the normal and superconducting sides. Such a
mismatch is well-known to act as an effective barrier for NIS
junctions. Consequently, additional barrier created by the gate
voltage becomes irrelevant, and we expect the tunneling conductance
to become independent of the applied gate voltage $V_0$. Also note
that at $eV =E_F$, $\alpha_c=0$ (Eq.\ \ref{criti1}). Hence there is
no Andreev reflection and consequently $G_0$ vanishes for all values
of the applied gate voltage for this bias voltage. Our results in
this limit, coincides with those of Ref.\ \onlinecite{beenakker1}.
Finally in Fig.\ \ref{figm6}, we show the dependence of amplitude of
oscillation of zero-bias tunneling conductance on $U_0$ for the
applied bias voltages $V_0 =6E_F$ and $\Delta_0=0.01 E_F$. As
expected, the oscillation amplitude with decreases monotonically
with increasing $U_0$. We have verified that this feature is
independent of the applied gate voltage $V_0$ as long as $V_0 \ge
V_{0c}$.

\section{Experiments}
\label{experiments}
\begin{figure}
\rotatebox{0}{
\includegraphics*[width=\linewidth]{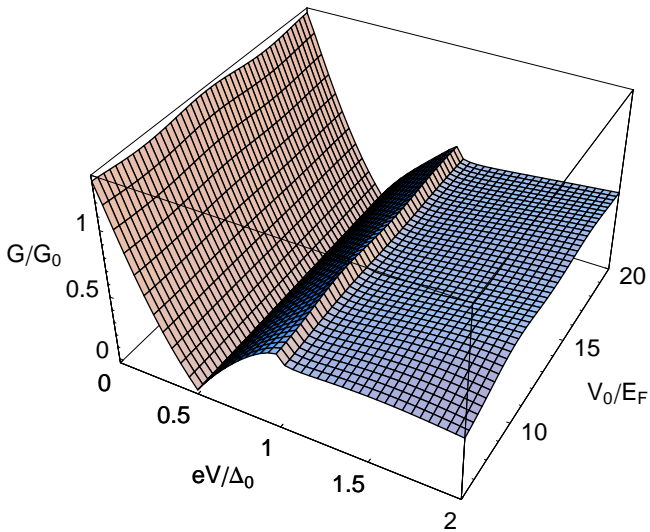}}
\caption{(Color online) Plot of tunneling conductance as a function
of the bias-voltage $eV$ and the gate voltage $V_0$ for $d=0.4
\lambda$, $\Delta_0=2E_F$ and $U_0 =25 E_F$. As discussed in the
text, the tunneling conductance is virtually independent of the
applied gate voltage $V_0$ due to the presence of a large $U_0$.
Note that maximum angle of incidence for which Andreev reflection
can take place vanishes at $eV = E_F$ leading to vanishing of $G$ at
this bias voltage.} \label{figm5}
\end{figure}
Superconductivity has recently been experimentally realized in
graphene \cite{delft1}. In our proposed experiment, one needs to
realize an NIS junction in graphene. The local barrier can be
fabricated using methods of Ref.\ \onlinecite{nov2}. The easiest
experimentally achievable regime corresponds to $\Delta_0 \ll E_F$
with aligned Fermi surfaces for the normal and superconducting
regions. We suggest measurement of tunneling conductance curves at
zero-bias ($eV=0$) in this regime. Our prediction is that the
zero-bias conductance will show an oscillatory behavior with the
bias voltage. In graphene, typical Fermi energy can be $E_F \le
40$meV and the Fermi-wavelength is $\lambda \geq 100$nm
\cite{geim1,nov2,delft1}. Effective barrier strengths of $\le 80$meV
\cite{geim1} and barrier widths of $d \simeq 10-50$ nm therefore
specifies the range of experimentally feasible junctions
\cite{geim1,nov2}. Consequently for experimental junctions, the
ratio $V_0/E_F$ can be arbitrarily large within these parameter
ranges by fixing $V_0$ and lowering $E_F$. Experimentally, one can
set $5 \le E_F \le 20$meV so that the conditions $\Delta_0 \ll E_F$
$V_0/E_F \gg 1$ is easily satisfied for realistic $\Delta_0 \sim
0.5$meV and $V_0=200$meV. This sets the approximate range $V_0/E_F
\ge 10$ for the experiments. Note that since the period (amplitude)
of oscillations increases (decreases) as $V_0/E_F \to 1$, it is
preferable to have sufficiently large values of $V_0/E_F$ for
experimental detection of these oscillations.

To check the oscillatory behavior of the zero-bias tunneling
conductance, it would be necessary to change $V_0$ in small steps
$\delta V_0$. For barriers of a fixed width, for example with values
of $d/\lambda=0.3$, it will be enough to change $V_0$ in steps of
approximately $20-30$meV, which should be experimentally feasible.
We note that for the above-mentioned range of $V_0/E_F$, the
experimental junctions shall not always be in the thin barrier
limit. For example, as is clear from Fig.\ \ref{figm2}, the
periodicity of oscillations $\chi_{\rm period}$ of the zero-bias
tunneling conductance of such junctions shall be a function of $V_0$
and shall differ from $\pi$. This justifies our theoretical study of
NIS junctions in graphene which are away from the thin barrier
limit.

Apart from the above-mentioned experiments, it should also be
possible to measure the tunneling conductance as a function of the
applied bias voltage $eV/ \Delta_0$ for different applied gate
voltages $V_0$. Such measurements can be directly compared with
Fig.\ \ref{figm3}. Finally, it might be also possible to create a
relative bias $U_0$ between the Fermi surfaces in the normal and
superconducting side and compare the dependence of oscillation
amplitudes of zero-bias tunneling conductance on $U_0$ with the
theoretical result shown in Fig.\ \ref{figm5}.

In conclusion, we have presented a theory of tunneling conductance
of graphene NIS junctions with barriers of thickness $d \ll \xi$ and
arbitrary gate voltages $V_0$ applied across the barrier region. We
have demonstrated that the oscillatory behavior of the tunneling
conductance, previously derived in Ref.\ \onlinecite{bhattacharya1}
for junctions with thin barriers, persists for all such junctions.
However, the periodicity and amplitude of these oscillations deviate
from their universal values in the thin barrier limit and become
functions of the applied barrier voltage $V_0$. We have also shown
that our work, which extends the earlier results of Ref.\
\onlinecite{bhattacharya1}, correctly reproduce the earlier results
for tunneling conductance obtained for thin \cite{bhattacharya1} and
zero \cite{beenakker1} barriers as limiting cases. We have discussed
experimental relevance of our results.

KS and SB thank Graduate Associateship Program at Saha Institute
which made this work possible. SB thanks T. Senthil and V.B. Shenoy
for stimulating discussions.


\begin{thebibliography}{99}
\bibitem{comment1} Since our solution for the wavefunction at the
barrier region breaks down at $V_0=E_F$, it is necessary to
carefully analyze the limit $V_0 \to E_F$ for arriving at this
conclusion.
\bibitem{comment2} Note that at $V_0=E_F$, $\alpha_c=0$ leading to $G=0$.
Thus the divergence of periodicity do not constitute a physical
singularity.
\end{thebibliography}
\end{document}